\documentclass[12pt, a4paper]{amsart}

\usepackage[utf8]{inputenc}
\usepackage[english]{babel}
\usepackage{lmodern}
\usepackage{textcomp}
\usepackage{extarrows}
\usepackage{amsmath,amssymb,amsthm}
\usepackage{comment}
\usepackage{geometry}
\usepackage{setspace}
\usepackage[dvipsnames]{xcolor}
\definecolor{cornellred}{rgb}{0.7, 0.11, 0.11}
\usepackage[colorlinks=true,urlcolor=RoyalBlue,citecolor=RoyalBlue,linkcolor=RoyalBlue,linktocpage,pdfpagelabels,
bookmarksnumbered,bookmarksopen]{hyperref}

\usepackage[round,authoryear]{natbib}

\usepackage{mathtools}
\mathtoolsset{showonlyrefs}
\usepackage{chemarrow}

\usepackage{tikz}
\usetikzlibrary{arrows.meta, decorations.pathmorphing, patterns, calc, positioning}
\usetikzlibrary{decorations.pathmorphing,shapes}

\newcounter{sarrow}
\newcommand\xrsquigarrow[1]{%
\stepcounter{sarrow}%
\begin{tikzpicture}[decoration=snake]
\node (\thesarrow) {\strut#1};
\draw[->,decorate] (\thesarrow.south west) -- (\thesarrow.south east);
\end{tikzpicture}%
}
\usepackage{caption}
\usepackage{csquotes}
\usepackage{booktabs}
\usepackage{array}
\usepackage{xcolor}
\usepackage{fancyhdr}

\title[Becoming, Duration, and the Evolving Block Universe]{Becoming, Duration, and the Evolving Block Universe}

\author[N.\,Cangiotti]{Nicol\`o Cangiotti}

\address[N.\,Cangiotti]{ISIS Paolino d'Aquileia\newline\indent
 Via Istituto Tecnico Agrario, 42\newline\indent 33043 Cividale del Friuli, Udine, Italy}
\email{canjioh@gmail.com}

\date{}

\begin{document}
\bibliographystyle{plainnat}

\begin{abstract}
\noindent
The two-times problem asks how the directed, oriented time of biological and conscious systems relates to the time of fundamental physics. The Evolving Block Universe, in which spacetime grows through an irreversible quantum dynamics along physically preferred worldlines, accounts for the direction of physical time and for its emergence across scales up to the time kept by the brain. It leaves open the relation between this physical time and the experience of passage. We argue that Bergson's durée, the continuous qualitative experience of lived time, names this residue and marks the first-person level at which the physical hierarchy reaches its limit. On this reading the two accounts divide the problem between them: the Evolving Block Universe describes its structural, third-person half, and durée its experiential, first-person half. The remaining gap between the two halves is explanatory. It is an instance of the familiar gap between the physical description of a system and the character of its states as they are lived, and it is compatible with physicalism.
\end{abstract}

\maketitle
\enlargethispage{-1.5cm}


\section{Introduction}
\label{sec:intro}

A current discussion in the foundations of physics concerns what \citet{BuonomanoRovelli2023} call the \emph{two-times problem}: the apparent mismatch between the time of fundamental physics, which is direction-neutral and has no privileged present, and the time of biological and cognitive systems, which is oriented, has a moving present, and underwrites memory, anticipation, and agency. \citet{BuonomanoRovelli2023} seek to bridge the gap through a closer integration of neuroscience and physics, conceding to physics the absence of a fundamental present and to neuroscience the centrality of oriented experience. \citet{GruberMontemayorBlock2020} propose a dualistic model built on the notion of an Information Gathering and Utilizing System (IGUS), in which veridical and illusory components of manifest time are distinguished. \citet{Ellis2024} locates the source of the difficulty in the static Block Universe (BU) itself and replaces it with a different physical model of spacetime, the Evolving Block Universe (EBU), in which the manifold grows along physically preferred worldlines through an irreversible quantum dynamics \citep{Ellis2006, Ellis2014}. Our proposal is that Ellis's EBU, read in conjunction with Bergson's philosophy of time, offers a more satisfactory treatment of the two-times problem than any of these programmes taken singly, because it keeps the structural and the experiential halves of the problem distinct.

The EBU is among the most developed of these proposals at the level of physics, and it is the one we take up here. Even granting its physical structure in full, however, the EBU leaves the two-times problem partly open, since it does not specify how the objectively evolving time of the growing manifold is related to the time we experience as passing. The experiential side of the problem has been pursued separately, by neuroscientific means, most fully in the dualistic continuity-illusion model of \citet{GruberMontemayorBlock2020, GruberBlockMontemayor2022} and \citet{MontemayorGruber2025}. In the terms we shall adopt, the EBU answers the structural, third-person half of the two-times problem, and the experiential, first-person half remains to be addressed. We argue that this second half can be clarified by recovering some of the insights of Bergson's philosophy of time. In his 1922 debate with Einstein, Bergson held that the time of special relativity is a \emph{measure} of time, obtained by abstraction from the qualitatively heterogeneous and continuous time of lived experience, which he called \emph{dur\'ee}; Einstein replied that the philosopher's time has no place in physics, and the subsequent consensus sided with Einstein \citep{Canales2015}. For a recent reassessment of that debate, see the essays collected in \citet{CampoGozzano2021}.

Ellis stratifies physical time into a hierarchy of parameters, from cosmological proper time $\tau_{\mathrm{global}}$ through local proper time $\tau_{\mathrm{local}}$ to brain time $T_B$, and observes that $T_B$ is not linearly related to proper time \citep{Ellis2024}. This non-linearity is itself a third-person fact, since it concerns the way in which the brain, considered as a physical clock, coarse-grains proper time; its effect on the two-times problem is to relocate the gap one level down, to the relation between $T_B$ and the passage of time as it is lived. Our claim is that \emph{dur\'ee} names this first-person level and completes the hierarchy as $\tau_{\mathrm{global}} \to \tau_{\mathrm{local}} \to T_B \to \textit{dur\'ee}$. The final step is of a different kind from the preceding ones: it marks the passage from a third-person to a first-person description of one and the same temporal process.

The contribution of the paper can now be stated. It offers an account of what any solution to the two-times problem must leave over, and of where the residue is located. The exchange between the EBU and Bergson runs in both directions and amounts to a division of labour. In one direction, the EBU gives physical content to three theses that Bergson could support only on phenomenological grounds, viz., that temporal irreversibility is structural, that physical time is inhomogeneous, and that time is the medium in which novelty is produced; each thesis has a counterpart in the formal structure of the EBU (Section~\ref{sec:convergence}). In the other direction, Bergson supplies the EBU programme with a vocabulary for the residual gap between physical and experiential time. We argue that this gap is the explanatory gap familiar from the philosophy of mind \citep{Levine1983}, which concerns the relation between two descriptions and is compatible with physicalism. This also distinguishes the proposal from the neuroscientific treatment of the same problem. The dualistic model assigns the felt passage of time to an adaptive and partly illusory system; we treat the felt passage as a first-person datum and argue that the limit it marks lies beyond the reach of any third-person description (Section~\ref{sec:spatialization}).

The paper is organised as follows. Section~\ref{sec:bu} formulates the static BU as a $B$-series picture in the sense of \citet{McTaggart1908}, recalls Merricks's objection to the Growing Block as a point of contrast, and states Ellis's reason for moving to a dynamical model. Section~\ref{sec:ebu} presents the EBU at the level of physical detail required here: Ricci eigenlines, proper time and the present surface, the ADM formalism, and wave-function collapse as the mechanism of irreversibility. Section~\ref{sec:hierarchy} develops the multi-scale hierarchy, distinguishes the Direction of Time from the local arrows of time, and returns to the two-times problem in order to locate the residual gap. Section~\ref{sec:bergson} reconstructs Bergson's critique of the spatialization of time. Section~\ref{sec:convergence} sets out the two-directional exchange between the EBU and \emph{dur\'ee}, characterises it as a structural homology, and explains why the EBU, among the available accounts of physical passage, is the appropriate counterpart of \emph{dur\'ee}. Section~\ref{sec:spatialization} answers the objection that the EBU's proper time is itself a spatialization of time, specifies the sense in which the residual gap is explanatory, and sets the resulting position against the dualistic continuity-illusion model of \citet{MontemayorGruber2025}.

\section{The static Block Universe as a \texorpdfstring{$B$}{B}-series picture}
\label{sec:bu}

In the standard Block Universe picture, spacetime is a four-dimensional, time-orientable Lorentzian manifold $(\mathcal{M}, g_{ab})$ that contains, as a single and unchanging whole, all the events that ever occur. The dynamical idea underwriting this picture is that, for globally hyperbolic spacetimes, initial data on a Cauchy surface $\Sigma$ determine a unique maximal development \citep[ch.~7]{HawkingEllis1973}, so that the entire history of the universe is fixed by a time-reversible evolution,
\begin{equation}
    H\colon S(\mathbf{x}, t_0) \longmapsto S(\mathbf{x}, t_1)\,,
    \quad \text{for all } t_1\,.
    \label{eq:hamiltonian}
\end{equation}
Since general covariance privileges no foliation of $\mathcal{M}$ and no value of the time coordinate $t$, the static BU contains no objective present, no temporal direction independent of boundary conditions or coarse-graining, and no becoming. Weyl's formulation of the view has become canonical: \enquote{the objective world simply is, it does not happen} \citep[cited in][]{BuonomanoRovelli2023}.

In the vocabulary of \citet{McTaggart1908}, the static BU corresponds to a $B$-series picture. Events are ordered by the tenseless relation \enquote{earlier than}, which in a time-oriented relativistic spacetime is naturally identified with the causal order $\preceq$, a partial order under which spacelike-separated events are incomparable; tensed properties such as being \emph{past}, \emph{present}, or \emph{future} are indexical descriptions adopted by observers located within the manifold. The position is usually called \emph{eternalism}, and it is a familiar reading of the spacetime formalism of relativity \citep[see, e.g.,][]{Hartle2005}. The reading is contested: \citet{Rovelli2019} defends a third option beyond presentism and eternalism, on which becoming is real but local and unoriented, a view the EBU shares except as regards the orientation of becoming (Section~\ref{sec:hierarchy}). A natural attempt to recover an $A$-series structure within a relativistic setting is the Growing Block Universe (GBU) of \citet{Broad1923}, in which past and present exist, the future does not, and new \enquote{layers} of reality are added at the leading edge. \citet{Merricks2006} objects that observers located in past layers cannot, on this proposal, distinguish themselves from observers located at the leading edge: from the inside the past appears as the present, and the proposal supplies no internal criterion by which an observer could establish that they occupy the privileged edge.

Merricks's argument clarifies what a successful refinement of the Block Universe picture has to provide, namely a physical structure internal to the manifold that distinguishes the leading edge from earlier layers. The EBU of \citet{Ellis2014} supplies a structure of this kind: its leading edge is the present surface $\mathcal{S}(\tau)$, defined by integrating proper time along the Ricci eigenlines (Section~\ref{sec:ebu}). 

A second consideration points in the same direction. \citet{Ellis2024} emphasises that the static BU, taken as a complete physical description, leaves no theoretical room for the directedness that biological and cognitive systems display. The objection concerns the conditions under which an appearance of passage could arise at all. Any framework that treats passage as an appearance must supply a substrate in which the appearance is instantiated, and the static BU, which is indifferent to the value of $t$ that is present, supplies none. In Ellis's words, \enquote{you can't have any experience whatever if the passage of time is an illusion: there is then no substrate that will make any illusion of any kind possible} \citep{Ellis2024}.

\section{The Evolving Block Universe: physical structure}
\label{sec:ebu}

In the passage from special to general relativity, the metric tensor $g_{ab}(\mathbf{x})$ is determined by the matter content of the universe through the Einstein field equations, written here in geometrised units ($G = c = 1$),
\begin{equation}
    G_{ab} = 8\pi T_{ab}\,.
    \label{eq:efe}
\end{equation}
Because no perfect vacuum exists in the real universe (the cosmic microwave background, intergalactic plasma, and dark sectors all contribute to a non-vanishing $T_{ab}$ on cosmological scales), Minkowski spacetime does not physically exist, and the global Lorentz symmetry that distinguishes the special-relativistic setting is broken by the actual matter distribution. This allows the definition of \emph{preferred timelike worldlines}: the eigenlines $x^a(v)$ of the Ricci tensor $R_{ab}$, whose 4-velocities $u^a(v) = dx^a(v)/dv$ satisfy
\begin{equation}
    T^{a}{}_{b}\, u^b = \lambda_1\, u^a
    \quad \Longleftrightarrow \quad
    R^{a}{}_{b}\, u^b = \lambda_2\, u^a,
    \label{eq:ricci_eigen}
\end{equation}
where the equivalence follows from~\eqref{eq:efe} in its trace-reversed form, $R_{ab} = 8\pi\,(T_{ab} - \tfrac{1}{2}\,T\,g_{ab})$, with $T = g^{ab}T_{ab}$ and $\lambda_2 = 8\pi(\lambda_1 - \tfrac{1}{2}T)$. These lines represent the average local motion of matter. For realistic matter, whose energy-momentum tensor is of Hawking--Ellis type~I, $T^{a}{}_{b}$ has a unique timelike eigenvector up to normalisation \citep[ch.~4]{HawkingEllis1973}, and the eigenlines are therefore geometrically unique \citep{Ellis2014}. The preferred worldlines are thus fixed by the physical content of the manifold through Einstein's equations.

\begin{figure}[ht]
\centering
\begin{tikzpicture}[scale=0.8, transform shape, yshift=0.6cm]

  \definecolor{pastblue}{RGB}{180,210,240}
  \definecolor{presentyellow}{RGB}{255,220,80}
  \definecolor{worldred}{RGB}{200,60,60}

  \def\bxa{0}
  \def\bya{0}
  \def\bxb{4}
  \def\byb{0}
  \def\height{3.5}
  \def\depth{1.2}
  \def\depthangle{30}

  \pgfmathsetmacro{\dx}{\depth*cos(\depthangle)}
  \pgfmathsetmacro{\dy}{\depth*sin(\depthangle)}

  \begin{scope}[xshift=0cm]
    \fill[pastblue, opacity=0.7]
      (\bxa,\bya) -- (\bxb,\byb) -- (\bxb,\byb+\height) -- (\bxa,\bya+\height) -- cycle;
    \fill[presentyellow!70, opacity=0.75]
      (\bxa,\bya+\height) -- (\bxb,\byb+\height)
      -- (\bxb+\dx,\byb+\height+\dy) -- (\bxa+\dx,\bya+\height+\dy) -- cycle;
    \fill[pastblue!40, opacity=0.6]
      (\bxb,\byb) -- (\bxb+\dx,\byb+\dy)
      -- (\bxb+\dx,\byb+\height+\dy) -- (\bxb,\byb+\height) -- cycle;
    \draw[thick]
      (\bxa,\bya) -- (\bxb,\byb) -- (\bxb,\byb+\height) -- (\bxa,\bya+\height) -- cycle;
    \draw[thick]
      (\bxa,\bya+\height) -- (\bxa+\dx,\bya+\height+\dy) -- (\bxb+\dx,\byb+\height+\dy);
    \draw[thick]
      (\bxb,\byb+\height) -- (\bxb+\dx,\byb+\height+\dy);
    \draw[thick,dashed, gray]
      (\bxa,\bya) -- (\bxa+\dx,\bya+\dy) -- (\bxb+\dx,\byb+\dy);
    \draw[thick,dashed, gray]
      (\bxa+\dx,\bya+\dy) -- (\bxa+\dx,\bya+\height+\dy);
    \draw[thick]
      (\bxb,\byb) -- (\bxb+\dx,\byb+\dy);
    \draw[thick]
      (\bxb+\dx,\byb+\dy) -- (\bxb+\dx,\byb+\height+\dy);
    \draw[very thick, color=presentyellow!80!black]
      (\bxa,\bya+\height) -- (\bxb,\byb+\height)
      -- (\bxb+\dx,\byb+\height+\dy) -- (\bxa+\dx,\bya+\height+\dy) -- cycle;
    \draw[-{Stealth[length=8pt]}, very thick]
      (\bxa-0.5,\bya) -- (\bxa-0.5,\bya+\height+0.6) node[left] {$\tau$};
    \node at (\bxa+1.0,\bya+\height/2) {\large $\mathcal{M}(\tau_1)$};
    \node[fill=presentyellow, rounded corners=2pt, inner sep=2pt, font=\small]
      at (\bxa+2.0,\bya+\height+\dy+0.45)
      {$\mathcal{S}(\tau_1)$: \textit{present}};
    \draw[worldred, very thick]
      (\bxa+1.8,\bya+0.2)
      .. controls (\bxa+2.0,\bya+1.5) and (\bxa+1.6,\bya+2.5)
      .. (\bxa+1.8,\bya+\height);
    \node[worldred, right, font=\small]
      at (\bxa+2.0,\bya+\height/2-0.3) {worldline};
    \node[below] at (2+\dx/2,-0.45) {\textbf{(a)} $\mathcal{M}(\tau_1)$};
  \end{scope}

  \begin{scope}[xshift=7.5cm]
    \def\heightB{5.0}
    \fill[pastblue, opacity=0.7]
      (\bxa,\bya) -- (\bxb,\byb) -- (\bxb,\byb+\heightB) -- (\bxa,\bya+\heightB) -- cycle;
    \fill[presentyellow!70, opacity=0.75]
      (\bxa,\bya+\heightB) -- (\bxb,\byb+\heightB)
      -- (\bxb+\dx,\byb+\heightB+\dy) -- (\bxa+\dx,\bya+\heightB+\dy) -- cycle;
    \fill[pastblue!40, opacity=0.6]
      (\bxb,\byb) -- (\bxb+\dx,\byb+\dy)
      -- (\bxb+\dx,\byb+\heightB+\dy) -- (\bxb,\byb+\heightB) -- cycle;
    \fill[pastblue!90, opacity=0.35]
      (\bxa,\bya+\height) -- (\bxb,\byb+\height)
      -- (\bxb,\byb+\heightB) -- (\bxa,\bya+\heightB) -- cycle;
    \draw[thick]
      (\bxa,\bya) -- (\bxb,\byb) -- (\bxb,\byb+\heightB) -- (\bxa,\bya+\heightB) -- cycle;
    \draw[thick]
      (\bxa,\bya+\heightB) -- (\bxa+\dx,\bya+\heightB+\dy) -- (\bxb+\dx,\byb+\heightB+\dy);
    \draw[thick]
      (\bxb,\byb+\heightB) -- (\bxb+\dx,\byb+\heightB+\dy);
    \draw[thick,dashed, gray]
      (\bxa+\dx,\bya+\heightB+\dy) -- (\bxb+\dx,\byb+\heightB+\dy);
    \draw[thick,dashed, gray]
      (\bxa,\bya) -- (\bxa+\dx,\bya+\dy) -- (\bxb+\dx,\byb+\dy);
    \draw[thick,dashed, gray]
      (\bxa+\dx,\bya+\dy) -- (\bxa+\dx,\bya+\heightB+\dy);
    \draw[thick]
      (\bxb,\byb) -- (\bxb+\dx,\byb+\dy);
    \draw[thick]
      (\bxb+\dx,\byb+\dy) -- (\bxb+\dx,\byb+\heightB+\dy);
    \draw[dashed, gray, thick]
      (\bxa,\bya+\height) -- (\bxb,\byb+\height)
      -- (\bxb+\dx,\byb+\height+\dy) -- (\bxa+\dx,\bya+\height+\dy) -- cycle;
    \draw[very thick, color=presentyellow!80!black]
      (\bxa,\bya+\heightB) -- (\bxb,\byb+\heightB)
      -- (\bxb+\dx,\byb+\heightB+\dy) -- (\bxa+\dx,\bya+\heightB+\dy) -- cycle;
    \draw[-{Stealth[length=10pt]}, very thick, blue!80!black, line width=1.8pt]
      (\bxb+\dx+0.35,\bya+\height+0.02) -- (\bxb+\dx+0.35,\bya+\heightB-0.02)
      node[midway,right=4pt,font=\small,black] {growth $\Delta\tau$};
    \node at (\bxa+1.0,\bya+\heightB/2) {\large $\mathcal{M}(\tau_2)$};
    \node[fill=presentyellow, rounded corners=2pt, inner sep=2pt, font=\small]
      at (\bxa+2.0,\bya+\heightB+\dy+0.45)
      {$\mathcal{S}(\tau_2)$: \textit{present}};
    \node[gray, font=\small, right]
      at (\bxb+\dx+0.15,\bya+\height/2+\dy/2)
      {$\mathcal{S}(\tau_1)$ (now past)};
    \draw[worldred, very thick]
      (\bxa+1.8,\bya+0.2)
      .. controls (\bxa+2.0,\bya+2.0) and (\bxa+1.6,\bya+3.8)
      .. (\bxa+1.8,\bya+\heightB);
    \draw[-{Stealth[length=8pt]}, very thick]
      (\bxa-0.5,\bya) -- (\bxa-0.5,\bya+\heightB+0.6) node[left] {$\tau$};
    \node[below] at (2+\dx/2,-0.45)
      {\textbf{(b)} $\mathcal{M}(\tau_2)$,\; $\tau_2>\tau_1$};
  \end{scope}

  \draw[-{Stealth[length=10pt]}, thick]
    (4.8,7) -- (7.3,7)
    node[midway,above,font=\small] {isometric embedding};

\end{tikzpicture}

\caption{The Evolving Block Universe.
\textbf{(a)}~The universe at parameter value $\tau_1$: the spacetime manifold $\mathcal{M}(\tau_1)$ (blue region) is bounded to the future by the present surface $\mathcal{S}(\tau_1)$ (yellow boundary). The red curve is a fundamental worldline (Ricci eigenline). \textbf{(b)}~The universe at a later parameter value $\tau_2 > \tau_1$: the manifold has grown by the amount $\Delta\tau = \tau_2 - \tau_1$. The earlier manifold $\mathcal{M}(\tau_1)$ embeds isometrically into $\mathcal{M}(\tau_2)$, establishing the global Direction of Time. The former present $\mathcal{S}(\tau_1)$ is now part of the fixed past, and the region $\tau(p) > \tau_2$ is not part of $\mathcal{M}(\tau_2)$.}
\label{fig:ebu}
\end{figure}
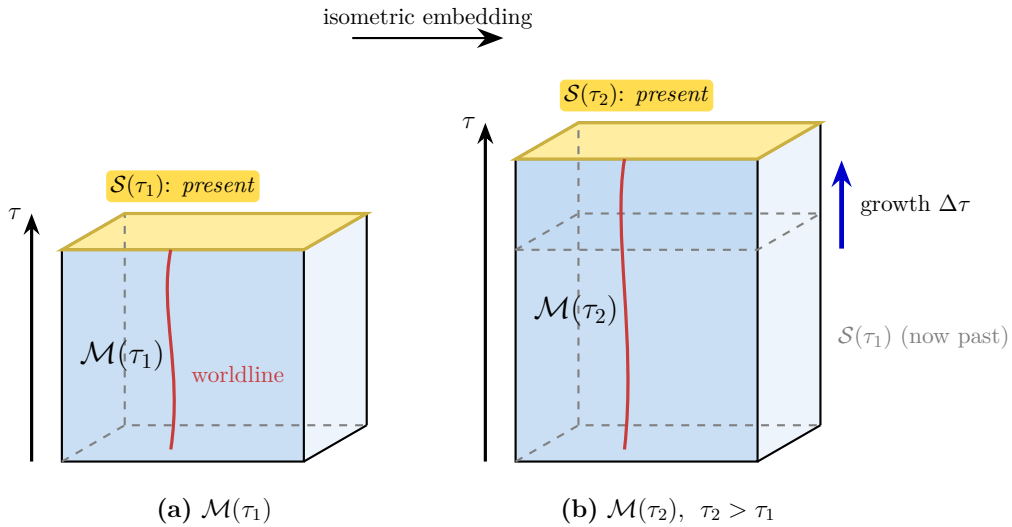

Proper time along these fundamental worldlines is determined by the metric in the usual way:
\begin{equation}
    \tau = \int \left(-g_{ab}\,\frac{dx^a}{dv}\frac{dx^b}{dv}\right)^{1/2}
    dv\,.
    \label{eq:proper_time}
\end{equation}
The present at parameter value $\tau_0$ is the spacelike surface
\begin{equation}
    \mathcal{S}(\tau_0) = \{\, p \in \mathcal{M} : \tau(p) = \tau_0\,\}\,,
    \label{eq:present_surface}
\end{equation}
and the EBU at parameter value $\tau_0$ is the spacetime manifold-with-boundary
\begin{equation}
    \mathcal{M}(\tau_0) = \{\, p \in \mathcal{M} : \tau(p) \leq \tau_0\,\}\,.
    \label{eq:ebu}
\end{equation}
With $\tau$ measured from a common initial surface (in the cosmological case, the initial singularity), the level sets~\eqref{eq:present_surface} are spacelike hypersurfaces wherever the gradient $\nabla_a\tau$ is timelike, as it is in the Friedmann--Lema\^{\i}tre--Robertson--Walker (FLRW) models, in which the comoving flow is geodesic and irrotational and $\tau$ coincides with cosmic time \citep[ch.~5]{HawkingEllis1973}. The hypersurfaces $\mathcal{S}(\tau_0)$ are then geometrically distinguished \enquote{present} slices, fixed by the matter content through~\eqref{eq:ricci_eigen}, and the freedom in the choice of foliation that characterises the static picture is removed. Earlier states are uniquely embedded into later ones through the isometric inclusion
\begin{equation}
    \mathcal{M}(\tau_1) \hookrightarrow \mathcal{M}(\tau_2)
    \quad \text{for } \tau_1 < \tau_2\,,
    \label{eq:embedding}
\end{equation}
so that the family $\{\mathcal{M}(\tau)\}_\tau$ is strictly increasing with respect to inclusion and spacetime grows monotonically toward the future (see Figure~\ref{fig:ebu}). In this picture the past is the realised region of spacetime, and the region $\{p : \tau(p) > \tau_0\}$, which is not yet physically determined, is by construction excluded from $\mathcal{M}(\tau_0)$. The directed growth of $\mathcal{M}(\tau)$ along the inclusion~\eqref{eq:embedding} is what \citet{Ellis2024} calls the global \emph{Direction of Time}, to be distinguished from the various local arrows of time discussed in Section~\ref{sec:hierarchy}.

The dynamical evolution of the metric in the EBU is governed by the Arnowitt--Deser--Misner ($3+1$) decomposition \citep{ArnowittDeserMisner1962}, which expresses general relativity in a manifestly dynamical form by foliating spacetime into a one-parameter family of spatial hypersurfaces $\Sigma_t$. The line element takes the standard form
\begin{equation}
    ds^2 = \left(-N^2 + N_i N^i\right) dt^2
    + 2 N_i\, dx^i\, dt
    + g_{ij}\, dx^i\, dx^j\,,
    \label{eq:adm_metric}
\end{equation}
where $N$ is the lapse function and $N^i$ is the shift vector. The evolution equations for the spatial metric $g_{ij}$ and its conjugate momentum $\pi^{ij}$ are
\begin{align}
    \partial_t g_{ij}
    &= 2N g^{-1/2}
    \left(\pi_{ij} - \tfrac{1}{2}g_{ij}\pi\right)
    + N_{i|j} + N_{j|i}\,,
    \label{eq:adm_g}\\
    \partial_t \pi^{ij}
    &= -N\sqrt{g}\left(
    {}^{(3)}R^{ij} - \tfrac{1}{2}g^{ij}{}^{(3)}R
    \right) + \cdots\,,
    \label{eq:adm_pi}
\end{align}
subject to the Hamiltonian and momentum constraints
\begin{align}
    {}^{(3)}R + g^{-1}\left(\tfrac{1}{2}\pi^2 - \pi_{ij}\pi^{ij}\right) &= 16\pi \rho_H\,,
    \label{eq:constraint1}\\
    -2 \pi^{ij}{}_{|j} &= 16\pi T^{i0}\,,
    \label{eq:constraint2}
\end{align}
with $\rho_H = n_a n_b T^{ab}$ the matter-energy density measured by observers normal to $\Sigma_t$. In the standard ADM formulation the lapse $N$ and the shift $N^i$ are freely specifiable, and their choice fixes the foliation and the spatial coordinates. The EBU removes this freedom for the foliation. It identifies the physically distinguished slicing with the family $\{\mathcal{S}(\tau)\}_\tau$ defined by the Ricci eigenlines, so that, with $t = \tau$, the lapse is fixed as $N = (-g^{ab}\nabla_a\tau\,\nabla_b\tau)^{-1/2}$ and the shift retains only the freedom of choosing spatial coordinates on the slices; and it integrates the evolution equations~\eqref{eq:adm_g}--\eqref{eq:adm_pi} only up to the present value $\tau_0$, so that the evolved region is $\mathcal{M}(\tau_0)$.

The EBU requires, alongside the geometric ingredients above, a physical mechanism for the irreversibility of the growth~\eqref{eq:embedding}. The candidate offered by \citet{Ellis2014} is the standard quantum-mechanical wave-function collapse, treated as an objective physical process that turns indeterminate superpositions into determinate eigenstates. Within collapse-based interpretations of quantum mechanics, the transition takes the schematic form
\begin{equation}
    \psi_1(\mathbf{x}) = \sum_n c_n u_n(\mathbf{x})
    \;\;\xrightarrow{\;t = t^*\;}\;\;
    \psi_2(\mathbf{x}) = a_N\, u_N(\mathbf{x})\,,
    \label{eq:collapse}
\end{equation}
with the index $N$ selected with probability $|c_N|^2$ according to the Born rule, and $a_N = c_N/|c_N|$ a phase factor. The transition~\eqref{eq:collapse} has three features that are directly relevant to the EBU. First, it is \emph{non-unitary}. Writing $P_N = |u_N\rangle\langle u_N|$ for the rank-one projector onto the eigenstate $u_N$, the transition is the map $\psi \mapsto P_N\psi/\lVert P_N\psi\rVert$, which is non-linear and sends distinct initial states onto the same final state; being many-to-one, it is not injective, and no unitary operator, hence no Schrödinger evolution, represents it. Second, it is \emph{informationally irreversible}: the final state $\psi_2$ does not determine the initial superposition $\psi_1$, since the coefficients $\{c_n\}_{n \neq N}$ of the unrealised alternatives are erased, and the transition admits no exact retrodiction (a unitary evolution $U$ is inverted by $U^{\dagger}$, which recovers the initial state). Third, it is \emph{temporally asymmetric}: it maps a plurality of potential outcomes onto a single realised actuality and admits no inverse, so that the direction from $\psi_1$ to $\psi_2$ is physically distinguished from its reverse. Within the EBU framework this asymmetry belongs to the mechanism itself and holds for individual events, independently of coarse-graining, statistical approximation, or the incompleteness of microscopic knowledge. Quantum collapse thereby supplies a microphysical grounding for the Direction of Time that is independent of any statistical argument.

\citet{Ellis2014} is explicit that the collapse~\eqref{eq:collapse} extends beyond controlled laboratory experiments: it \enquote{happens all the time everywhere, it does not need to relate to an experiment} \citep{Ellis2014}. A radioactive nucleus decaying in a mineral grain, a photon absorbed by a rhodopsin molecule and triggering the isomerisation of its retinal chromophore, and a quantum fluctuation in the early universe seeding a future galaxy cluster are, on this picture, instances of the same asymmetric transition, and the block grows continuously and pervasively, whether or not an observer is present. The use of collapse in the EBU is conditional on the adoption of a collapse-based interpretation of quantum mechanics; decoherence-based and many-worlds interpretations would require a different mechanism to play the same role.

Ellis calls the process by which collapse events bring about the growth of the manifold \emph{concretizing} \citep{Ellis2014}: each collapse event concretizes one of the many potential futures into the single actual past. Each act of concretization produces a state of the manifold that the prior state does not uniquely determine. The outcome $\psi_2 = a_N u_N(\mathbf{x})$ is fixed by $\psi_1$ only up to the probabilities $\{|c_n|^2\}$, and it is realised through the interaction with a macroscopic context in an intrinsically unpredictable way, so that the post-concretization state of the universe cannot be deduced from its prior state. This is the counterpart, within the foundations of physics, of Bergson's claim that time is the medium in which novelty is produced, to which we return in Section~\ref{sec:convergence}.

\section{The multi-scale hierarchy and the Direction of Time}
\label{sec:hierarchy}

A central feature of the EBU that distinguishes it from the bare Growing Block proposal is its explicit multi-scale structure \citep{Ellis2014}. Ellis identifies five scales of physical description, from quantum gravity (Scale~0) and microphysics (Scale~1) through the macro-level of daily life and biology (Scale~2) to astrophysical structures (Scale~3) and cosmology (Scale~4); curvature is determined at Scale~3, and local physical events by the interaction of Scales~1 and~2. The stratification has a temporal significance: each scale carries its own temporal parameter, and higher parameters emerge from lower ones without reducing to them.

At the cosmological and astrophysical scales (Scales 3--4), the relevant temporal quantity is the global proper time $\tau_{\mathrm{global}}$ defined along Ricci eigenlines by~\eqref{eq:proper_time}, the parameter with respect to which the manifold $\mathcal{M}(\tau)$ grows. At Scale~2, the relevant parameter is the local proper time $\tau_{\mathrm{local}}$ along individual observer worldlines. For a worldline $\gamma$ parametrized by $\tau_{\mathrm{global}}$, the two are related by $d\tau_{\mathrm{local}}/d\tau_{\mathrm{global}} = (-g_{ab}\,\dot\gamma^a\dot\gamma^b)^{1/2}$, with $\dot\gamma^a = d\gamma^a/d\tau_{\mathrm{global}}$, so that $\tau_{\mathrm{local}}$ differs from $\tau_{\mathrm{global}}$ by the kinematic and gravitational time dilation accumulated along the path. At the intersection of Scales~2 and~1, \citet{Ellis2024} introduces a further parameter, brain time $T_B$, defined as the temporal parameter that governs the firing patterns of neural oscillators and the cognitive integration of temporal information. Ellis is explicit that \enquote{brain time $T_B$ and proper time $\tau$ are not linearly related} \citep{Ellis2024}: the map $\tau_{\mathrm{local}} \mapsto T_B$ is not affine, since the rate $dT_B/d\tau_{\mathrm{local}}$ varies with attention, emotional state, and the internal dynamics of multilevel oscillatory networks, and the brain thereby performs a contextual coarse-graining of local proper time. The three parameters form an ordered chain:
\begin{equation}
\begin{aligned}
\tau_{\mathrm{global}}&\;\text{(cosmological proper time, Scales 3--4)};
\\
\tau_{\mathrm{local}} &\;\text{(local proper time, Scale 2)};
\\
T_B &\;\text{(brain time, Scales 1--2)}.
\end{aligned}
\label{eq:tau_h}
\end{equation}
Each level plays its own causal role within its domain and carries physical information that is not obtainable from the level above. The non-linearity of $T_B$ as a function of $\tau_{\mathrm{local}}$ is, in particular, a physical fact about biological systems, measurable in principle by the methods of neuroscience.

A distinction that has not always been clearly drawn in the philosophy of time, and that is essential for what follows, is the distinction between the \emph{Direction of Time} and the \emph{arrows of time} \citep{Ellis2014, Ellis2024}. The Direction of Time is the global, structural asymmetry established by the EBU through the isometric inclusion~\eqref{eq:embedding}, that is, the fact that $\mathcal{M}(\tau_1)$ embeds uniquely and properly into $\mathcal{M}(\tau_2)$ for $\tau_1 < \tau_2$, with $\mathcal{S}(\tau)$ a moving spacelike boundary between the fixed past and the undetermined future. The arrows of time (thermodynamic, electromagnetic, gravitational-radiative, quantum) are local and derivative: they are specific physical processes that exhibit a preferred temporal sense, and they align with the global Direction of Time. Eddington's formulation of the thermodynamic arrow \citep{Eddington1928} requires special initial conditions in addition to the microphysical laws, which are invariant under time reversal. Boltzmann's $H$-theorem obtains the monotonic decrease of $H$ from the time-asymmetric assumption of molecular chaos, and the observed increase of entropy depends on special initial conditions within the cosmological context of an expanding universe \citep[see also][]{HawkingEllis1973}. In the EBU the time-reversal symmetry is broken by the growing boundary, and the local arrows align with it. On this point the EBU diverges most sharply from the deflationary account of temporal orientation defended by \citet{Rovelli2017, Rovelli2023}, on which the orientation of time and of causation is perspectival and rooted in the entropy gradient. On the EBU the orientation is structural, and it is carried by the growth of $\mathcal{M}(\tau)$.

The distinction bears on the comparison with Bergson developed in Section~\ref{sec:convergence}. Bergson's critique was never directed against a local arrow of time, and he raised no objection to the Second Law as a macroscopic generalisation. His target was the thesis that temporal asymmetry is purely statistical, an artefact of coarse-graining that a Laplacian demon with complete microphysical knowledge could in principle reverse. Against this thesis Bergson held that the irreversibility of \emph{dur\'ee} is absolute. The EBU renders a structural irreversibility of this kind physically describable at the level of the Direction of Time. The growth of $\mathcal{M}(\tau)$ is one-directional: the events concretized between $\tau_1$ and $\tau_2$ belong to the fixed past, in the sense that the information carried by the pre-collapse amplitudes $\{c_n\}_{n \neq N}$ in~\eqref{eq:collapse} has been irrecoverably erased. For Bergson and for the EBU alike, this irreversibility is primary.

A related refinement concerns the structure of the present surface $\mathcal{S}(\tau)$. In the basic EBU it is a sharp spacelike hypersurface; in the \emph{Crystallizing Block Universe} (CBU) of \citet{EllisRothman2010}, quantum delayed-choice effects produce a \emph{crinkly} present, different spatial regions concretizing at slightly different proper times, so that $\mathcal{S}(\tau)$ acquires a non-zero thickness on Scale~1. The boundary between the determinate past and the indeterminate future is thus itself a physically structured region. Its Scale-2 counterpart is the specious present \citep{Poppel1997, GruberBlockMontemayor2022}, the neural integration window of approximately 2--3 seconds within which before and after are experienced as a unity; we return to it in Section~\ref{sec:convergence}.

We can now return to the two-times problem in the sharper form that the EBU makes available. The hierarchy~\eqref{eq:tau_h} is a chain of physical, third-person parameters. It records how cosmological proper time is related to local proper time, and how local proper time is coarse-grained by the brain into $T_B$; Ellis's observation that $T_B$ is non-linearly related to $\tau$ \citep{Ellis2024} belongs to this third-person register, since it concerns the brain considered as a physical clock. The hierarchy contains no term for the time that a conscious system experiences as passing. The EBU therefore gives the two-times problem a definite location: within the EBU, the problem reappears as the question of how brain time $T_B$ is related to the felt present. This is the difficulty pressed against the growing-block strategy in the recent literature. \citet{Dorato2023}, crediting \citet{GruberBlockMontemayor2022}, observes that the locally growing block \enquote{seems to be unable to shed light on the two-times problem from an empirical viewpoint}, and the experiential side is left to separate, neuroscientific means \citep{MontemayorGruber2025}, a programme we discuss in Section~\ref{sec:spatialization}. One philosophical response, discussed in this connection by \citet{Dorato2023}, treats the growing block as a primitive \enquote{irreducible intrinsic asymmetry in the temporal structure of the universe} \citep[in the sense of][]{Maudlin2007}, and the EBU can be read as a physical realisation of such an asymmetry. The final term of the hierarchy, the time of experience, lies beyond the resources of the EBU, and we argue that Bergson's analysis of \emph{dur\'ee} supplies it.

\section{Bergson's \textit{dur\'ee} and the spatialization critique}
\label{sec:bergson}

The time of experience, which the EBU cannot reach with its uniformly third-person parameters, is Bergson's point of departure: his philosophy of time turns on the first-person character of lived duration. We reconstruct his position here, before setting it, in the next section, against the EBU's physical structure.

Bergson's central argument rests on a diagnosis of what goes wrong when physics represents time as a dimension of the spatiotemporal manifold. The argument proceeds in three moments.

The first moment concerns the nature of lived time. In the \emph{Essai sur les donn\'ees imm\'ediates de la conscience}, \citet{Bergson1889} argues that the time of consciousness is a continuous flow in which each moment retains the past and anticipates the future: the past is carried forward and interpenetrates the present, and the present is directed toward an open future. Bergson thereby rejects the representation of conscious time as a succession of homogeneous, discrete instants laid out along an axis. The structure he describes is qualitatively heterogeneous in a definite sense. Two points of a geometric line differ only in position, and a translation of the line carries one into the other; two moments of \emph{dur\'ee} differ in their qualitative character, that is, in what they have accumulated and in what they are becoming, and no transformation of this kind relates them.

The second moment concerns the relation between lived time and physical time. When we represent duration on an axis, measure it with clocks, or coordinate it across observers, we project it onto a homogeneous and reversible medium that is, in its logical structure, spatial; in modern terms, onto the real line with its translation and reflection symmetries. This operation, which Bergson calls \emph{spatialization}, is legitimate and practically indispensable, since it is the operation by which physics turns the lived flow of time into a quantity amenable to calculation. Its effect is to abstract from the qualitative character of duration, so that spatialization leaves a parameter, a measure from which the qualitative movement has been removed.

The third moment identifies the philosophical error, which occurs when the spatialized representation is taken for a complete description of what there is to know about time. The static Block Universe commits this error in its most extreme form. By representing all the events of the universe as co-present in a four-dimensional manifold, which can be surveyed \enquote{all at once} from a standpoint outside time, it constructs a descriptive framework that is in principle unable to register the asymmetry between past and future, between the determinate and the open. The framework thereby lacks the descriptive resources required to represent what the experience of time is tracking.

Four properties of \emph{dur\'ee} matter for the comparison with the EBU. First, it is qualitatively heterogeneous: no two of its moments are identical, and its division into homogeneous units destroys its nature, as the division of a melody into separate notes destroys the melodic character that exists only in their succession. Second, it is continuous in the sense of interpenetration: its moments flow into one another (\emph{s'interp\'en\'etrent}), each carrying the memory of what has passed and the anticipation of what is coming. Third, it is irreversible in an absolute sense, since a \enquote{reversal} of time would annihilate the qualitative content that has accumulated. Fourth, it is accessible only from within, as the structure of lived experience. Bergson calls the cognitive mode appropriate to this access \emph{intuition}, an \enquote{intellectual sympathy by which one places oneself within an object in order to coincide with what is unique in it and consequently inexpressible} \citep{Bergson1934}; so understood, intuition grasps the qualitative movement of duration from which analysis, which yields static representations, abstracts. The fourth property makes \emph{dur\'ee} irreducibly first-personal: no third-person parametric description captures it in full.

In \emph{Dur\'ee et simultan\'eit\'e}, \citet{Bergson1922} engaged directly with special relativity. He argued that the multiple times of the theory, the coordinate times of different inertial frames, are multiple \emph{measurements} of one underlying duration, and that there must be a single real time, the duration of the universe itself, of which the physicist's parametric times are projections. \citet{Capek1971} and, more recently, \citet{Canales2015} have argued that Bergson's position was more nuanced than Einstein allowed, and that the dismissal rested in part on a failure to separate two distinct claims. The first, that \emph{dur\'ee} is a structure of experience irreducible to any parametric representation, is unaffected by Einstein's argument, which concerns the physical definition of simultaneity. The second, that physical times are projections of a single richer temporal reality, is harder to sustain, although the EBU lends it some support, since the multi-scale hierarchy~\eqref{eq:tau_h} orders the physical temporal parameters as a sequence of progressively coarser descriptions.

\section{\textit{Dur\'ee} and the hierarchy of times in the EBU}
\label{sec:convergence}

This section sets the physical structure of the EBU against Bergson's analysis of \emph{dur\'ee}, with the aim of showing how the EBU gives physical content to claims that Bergson could support only on phenomenological grounds. By \enquote{physical content} we mean the following. The EBU supplies a physical structure homologous to the \emph{structural} features of \emph{dur\'ee}, viz., the openness of the future, the reality of becoming, irreversibility, and heterogeneity. The qualitative character of lived duration remains on the first-person side, and it is the subject of Section~\ref{sec:spatialization}. The relation we draw is accordingly a structural correspondence between a physical model and a phenomenological description, and the two descriptions remain distinct.

Three of Bergson's claims acquire a physical reading in the EBU. The first is that the future is open. In the EBU this openness is a property of the physical description: by~\eqref{eq:ebu}, $\mathcal{M}(\tau_0)$ contains no point $p$ with $\tau(p) > \tau_0$, and $\mathcal{S}(\tau_0)$ marks the edge of what has been determined. The second is that becoming is real. The EBU describes a manifold that grows as quantum events concretize, so that each static $B$-series picture, the region $\mathcal{M}(\tau_0)$ taken at a fixed value of $\tau_0$, is an abstraction from the process of growth. The third is that the direction of time belongs to the structure itself. In the EBU it is carried by the non-unitary collapse~\eqref{eq:collapse} and by the one-way inclusion~\eqref{eq:embedding}, whose structural irreversibility answers, as argued in Section~\ref{sec:hierarchy}, to Bergson's claim that the irreversibility of \emph{dur\'ee} is absolute.

The comparison requires the EBU in particular, and the reason can be made precise by considering two alternative accounts on which physical time passes. On a moving-spotlight view \citep{Skow2015}, the whole manifold $\mathcal{M}$ exists tenselessly, and passage consists in the motion of a privileged hypersurface $\Sigma_{t_0}$ of a time function $t$; at each stage the region $\{p \in \mathcal{M} : t(p) > t_0\}$ belongs to the ontology, and the future events it contains await illumination. Such a view reproduces several of the parallels drawn above, and its future is fixed. A static block supplemented with a detailed physical model of neural processing is in the same position: it can represent the brain's construction of a felt present within a history settled in advance. The EBU differs from both in the structure of~\eqref{eq:ebu}, since $\mathcal{M}(\tau_0)$ excludes the region beyond $\mathcal{S}(\tau_0)$, which is physically undetermined. The difference bears directly on the comparison with Bergson. The Bergsonian datum comprises the experience that time passes \citep{Torrengo2017} and the creative character of duration, its being the medium in which novelty appears (Section~\ref{sec:ebu}). A framework that fixes the future in advance can model the feeling of passage, and it provides no physical correlate for the openness at which that feeling is directed; in the EBU this openness belongs to the physical description. The contrast is sharpest with the deflationary view on which experience does not present passage at all \citep{Prosser2016}; on the account defended here, the phenomenology of passage answers to a physical process of becoming.

Two further parallels concern the texture of time. Bergson's principal charge against the physicists' time was its homogeneity: it is a bare axis on which moments differ only in position. The EBU's $\tau$ is inhomogeneous in the relevant sense, because the eigenline condition~\eqref{eq:ricci_eigen} ties each value of $\tau$ to a definite configuration of matter and curvature; Section~\ref{sec:spatialization} states this property in terms of the symmetries of $\tau$. Bergson, moreover, took physical time to be an abstraction obtained from lived time through successive idealizations, and the EBU reaches a comparable picture from the side of physics: its parameters form a graded hierarchy~\eqref{eq:tau_h}, in which each level is emergent from the one above and irreducible to it, and no level constitutes the single true time.

The parallel that matters most for the two-times problem concerns experienced time itself. The EBU treats experienced time as real: brain time $T_B$ is an emergent parameter, non-linearly related to proper time. Bergson goes further and makes lived duration primary, and the two accounts can agree that experienced time is a structured feature of reality while disagreeing about what is primary. This shared commitment allows us to read \emph{dur\'ee} as the term that completes the hierarchy of the EBU:
\begin{equation}
\boxed{\;
\tau_{\mathrm{global}}
\;\xrightarrow{\text{emergent, top-down\,}}\;
\tau_{\mathrm{local}}
\;\xrightarrow{\text{coarse-graining\,}}\;
T_B
\;}
\;\;\xrsquigarrow{\footnotesize first-person\,\,}\;\;
\textit{dur\'ee}
\label{eq:hierarchy}
\end{equation}

The box encloses the parametrized, third-person part of the hierarchy, and the wavy arrow marks the single step that leaves it, namely the passage to a level that no physical parameter captures. Each of the two arrows inside the box expresses a relation of emergent irreducibility: the target level is real and causally efficacious, and it cannot be derived from its source by a straightforward reduction. The transition from $\tau_{\mathrm{global}}$ to $\tau_{\mathrm{local}}$ is governed by time dilation and local curvature, since the proper time of each observer depends on the path traversed through the manifold. The transition from $\tau_{\mathrm{local}}$ to $T_B$ is the one explicitly thematised by \citet{Ellis2024}: the brain coarse-grains physical proper time non-linearly, and the dualistic IGUS model of \citet{GruberMontemayorBlock2020, GruberBlockMontemayor2022}, with its distinction between veridical and illusory components of manifest time, operates at this level. The final transition, from $T_B$ to \emph{dur\'ee}, is of a different kind. $T_B$ is in principle measurable by neuroscience, through neural oscillators, attention-dependent distortions, and the specious-present window, and \emph{dur\'ee} is what these processes are like from the inside, in the sense of \citet{Nagel1974}. $T_B$ and \emph{dur\'ee} are thus two descriptions of one and the same process, the first given from without as neural time and the second from within as lived duration. \emph{Dur\'ee} is accordingly the irreducibly first-person level of the hierarchy, the level at which time is lived by a subject; Section~\ref{sec:spatialization} specifies the sense in which no physical theory carries out the final transition.

The identification in~\eqref{eq:hierarchy} expresses a claim of complementarity, and it involves no reduction of \emph{dur\'ee} to a physical parameter. One and the same temporal structure, the directed, heterogeneous and irreversible character of biological time, is described from outside by physics and neuroscience and from inside by phenomenology. The two descriptions are mutually irreducible, they bear on a single causally integrated process, and neither is dispensable.

The preceding comparisons establish a relation that should be characterised explicitly. We call it a \emph{homology}: an abstract form is realised twice, once as a feature of the growing manifold and once as a feature of lived time. Four correspondences constitute it. The openness of \emph{dur\'ee} corresponds to the exclusion of the region $\{p : \tau(p) > \tau_0\}$ from $\mathcal{M}(\tau_0)$ in~\eqref{eq:ebu}; directed becoming corresponds to the family $\{\mathcal{M}(\tau)\}_\tau$, strictly increasing under the inclusion~\eqref{eq:embedding}; absolute irreversibility corresponds to the non-invertibility of the collapse map~\eqref{eq:collapse}; and heterogeneity corresponds to the eigenline condition~\eqref{eq:ricci_eigen}, through which each value of $\tau$ is tied to a definite configuration of matter and curvature. A homology in this sense is weaker than an identity, which would make physical becoming and lived duration the same, and weaker than a reduction, which would derive the second from the first. It is stronger than a loose analogy, because each correspondence pairs a phenomenological feature with a definite element of the formal construction of the EBU. The homology therefore underwrites the division of labour proposed in Section~\ref{sec:intro}: the features that Bergson isolated by reflection on experience are structural facts, which physics, suitably formulated, also records. The qualitative character of duration is the single feature without a physical homologue, and the gap discussed in Section~\ref{sec:spatialization} opens at that point.

A further point concerns agency. In a static Block Universe the entire future history of every worldline is geometrically fixed: the neural states of a deliberating agent, including the outcome of the deliberation, are inscribed in the block independently of anything the agent does, so that agency reduces to a functional description of a predetermined causal chain. Ellis regards this consequence as untenable and argues that only in the EBU can an IGUS have agency, \enquote{which they manifestly do} \citep{Ellis2024}. We do not adopt the strong claim that the EBU is the only setting in which agency can be accommodated. Compatibilists have long held that agency is consistent with a fixed future, and we do not undertake to settle that debate here; our claim is comparative. The EBU provides room for agency because the future is, within the framework, physically undetermined: the outcome of each concretization event is fixed by no prior state, and an agent's deliberation can be construed as a selection among possibilities that no prior physical description specifies uniquely. The openness of the future over which deliberation ranges is thus, in the EBU, a matter of physics in addition to being a matter of the agent's epistemic situation.\footnote{Here too the EBU differs from \citet{Rovelli2021}, who traces the time-asymmetry of agency to thermodynamic irreversibility and the growth of entropy; on the EBU it follows from the structural indeterminacy of the future.} The point connects directly with Bergson. The \emph{Essai} \citep{Bergson1889} is, in its central argument, a defence of free will against associationist determinism: determinism, on Bergson's diagnosis, takes the spatialized representation of an action, mapped onto a homogeneous and reversible medium and viewed from outside, for the action as lived from within, where each moment of deliberation is qualitatively unique and the future is open. Bergson could support this openness only on phenomenological grounds; the EBU supplies a physical setting in which the future is undetermined.\footnote{\citet{Capek1971}, writing before the EBU programme was developed, argued that a rehabilitation of Bergson against Einstein's dismissal required, in addition to phenomenological grounds, a physics of irreversible processes, which in his view thermodynamics could not supply, the Second Law being statistical and therefore compatible with the block picture. The EBU delivers much of the physics that \v{C}apek anticipated: its irreversibility is structural and is grounded, within collapse-based interpretations, in the non-unitary character of quantum measurement.}

The specious present is the closest point of structural contact between the EBU and \emph{dur\'ee}. \citet{GruberBlockMontemayor2022}, following \citet{Poppel1997}, report that the brain integrates temporal events within a window of approximately 2--3 seconds, within which before and after are experienced as a single present. The phenomenon is close to what Bergson calls the interpenetration of moments in \emph{dur\'ee}, and the specious present can be read as the empirical, Scale-2 manifestation of that structure at the level of $T_B$. The notion goes back to \citet{James1890}, and the debate it has generated is surveyed by \citet{Dainton2023}. A finite window of $\sim 2$--$3\,\mathrm{s}$ is consistent with the static BU, and the two pictures accommodate it in different ways. In a static block the specious present is a feature of the way the brain represents time, with no correlate in the temporal structure of the world, which is the reading favoured by deflationary accounts of the manifest image of time \citep{Callender2017}. In the EBU it has, in addition, a worldly correlate: in the CBU refinement the boundary of the realised region has a non-zero thickness at Scale~1, set by the timescales of concretization, and the integration window of $T_B$ is its counterpart at Scale~2. The crinkly present of the CBU thus supplies a physical analogue of Bergson's claim that \emph{dur\'ee} is heterogeneous both across subjects and within the experience of a single subject, who lives the same objective interval as longer or shorter according to attention and content \citep{Bergson1889}. The analogue relates two distinct levels, since the crinkly present operates at Scale~1 and \emph{dur\'ee} belongs to the first-person level, and it establishes that the non-uniformity Bergson took to be characteristic of lived time has a counterpart in the physical architecture of the EBU.

\section{The spatialization objection revisited}
\label{sec:spatialization}

A Bergsonian might press the following objection. The proper time defined in~\eqref{eq:proper_time} is a real-valued function $\tau\colon \mathcal{M} \to \mathbb{R}$: it is measurable and representable on an axis, and the real line on which it takes values is invariant under translations and under reflection. Its level sets $\mathcal{S}(\tau_0)$ partition the manifold in a way that is formally identical to the spatial slicings Bergson objected to in Einstein's construction. On this objection, the EBU commits the spatialization error with a more sophisticated parameter.

The objection has force against a parameter that is homogeneous and reversible, and the reason can be stated in terms of symmetries. In Bergson's sense, a parameter $t$ is spatial when the structure it orders is invariant under the translations $t \mapsto t + a$, so that each value differs from every other only in position, and under the reflection $t \mapsto -t$, so that the parameter carries no intrinsic direction. These are the symmetries of the real line regarded as a one-dimensional Euclidean space, and the representation of time by a parameter with these symmetries is the spatialization that Bergson criticised. The question is therefore whether the physical structure parametrized by the EBU's $\tau$ inherits the symmetries of $\mathbb{R}$, and it inherits neither. Translation invariance is broken in two ways. First, the eigenline condition~\eqref{eq:ricci_eigen} fixes $\tau$ through the matter content, so that distinct values of $\tau$ correspond to distinct physical states. In the FLRW models, where the Ricci eigenlines are the worldlines of the comoving fluid and $\tau$ is cosmic time, the continuity equation $\dot\rho = -3H(\rho + p)$, with Hubble rate $H > 0$ and $\rho + p > 0$, makes the energy density $\rho$ a strictly decreasing function of $\tau$ \citep[ch.~5]{HawkingEllis1973}. Second, at each stage of its growth the EBU singles out the value $\tau_0$ as the boundary of the realised region $\mathcal{M}(\tau_0)$, a structure that no translation of $\tau$ preserves. Reflection invariance is broken because the family $\{\mathcal{M}(\tau)\}_\tau$ is strictly increasing under the inclusion~\eqref{eq:embedding} and the reflection $\tau \mapsto -\tau$ reverses this order; at the microphysical level, the collapse map~\eqref{eq:collapse} is non-invertible, and no physical process returns the region gained between $\tau_1$ and $\tau_2$ to non-existence. The homogeneity and reflection symmetry of $\mathbb{R}$ are thus properties of the mathematical representation, which the physical structure breaks; in Bergson's terms, $\tau$ measures time and is distinct from it. The multi-scale structure strengthens the reply. The parameter targeted by the objection, $\tau_{\mathrm{global}}$, belongs to the cosmological scale, and the time relevant to experience is $T_B$, which depends non-linearly on $\tau$ and carries the contextual, attention-dependent variation that Bergson took to be characteristic of duration. Whatever force the spatialization charge retains applies to $\tau_{\mathrm{global}}$, and $T_B$, the level at which the question of lived duration arises, is unaffected by it.

This reply meets the objection on its own ground and leaves a harder question open, on which we think Bergson is right. $T_B$ is context-sensitive, and it remains a third-person quantity of the kind that neuroscience measures; \emph{dur\'ee} is the same structure as it is lived from within. The gap between the two is real, and we hold that it is \emph{explanatory}. It is the explanatory gap of \citet{Levine1983}: a complete third-person description of a system leaves unexplained what its states are like in the first person \citep{Nagel1974}, which is the difficulty that \citet{Chalmers1995} calls the hard problem of consciousness, and lived time is a further instance of it. The thesis admits a precise statement. Let $P$ denote the complete physical description of the system, including the value and the dynamics of $T_B$, and let $Q$ denote the first-person character of its duration. Our claim is that $Q$ cannot be derived from $P$ by any explanation available to physics or neuroscience. The claim concerns derivability and intelligibility, and it leaves open whether $P$ metaphysically necessitates $Q$. Physicalism affirms this necessitation, on the view that \emph{dur\'ee} supervenes on the neural states that realise $T_B$, and its rivals deny it; the claim is compatible with both positions, nothing argued here decides between them, and the proposal does not depend on the choice. This is the sense in which, as stated in Section~\ref{sec:convergence}, no physical theory carries out the step from $T_B$ to \emph{dur\'ee}: the step is explanatory, and it carries no commitment to dualism. The gap is accordingly the boundary between a physical description and the experience that the description concerns, and a refinement of physical theory leaves it in place. Locating the gap at the step from $T_B$ to \emph{dur\'ee} keeps the two halves of the two-times problem distinct.

Our account also differs from the most developed neuroscientific treatment of the same problem. \citet{MontemayorGruber2025}, building on the dualistic model of \citet{GruberMontemayorBlock2020, GruberBlockMontemayor2022}, propose to solve the two-times problem by dividing human time into two systems. The first is a veridical system that mirrors the physical parameters described by the EBU; the second is a system of cerebral \enquote{continuity illusions} that generates the felt flow of time, which the authors call \enquote{non-illusory} in a pragmatic sense, since agency and motor control depend on it. We accept the model as an account of the function of the felt flow. The assignment of the felt flow to an adaptive system, however, answers a functional question, and an explanation of why a continuity illusion was selected leaves open what the resulting passage is like from the inside. The model therefore relocates the residue: it sorts lived time into a veridical mirroring and an adaptive supplement, and the explanatory gap reappears between each of these and the experience of it. The difficulty can be put as a dilemma. Either the continuity-illusion system is a further piece of neural machinery, in which case it belongs on the third-person side together with $T_B$, and the question of what its operation is like from within is left where it was; or it is the felt passage of time itself, in which case its description as an \enquote{illusion} misclassifies the datum that was to be explained. On neither horn is the first-person residue removed, and \emph{dur\'ee} is the name we give to it.

The EBU and \emph{dur\'ee} are therefore complementary. The EBU answers the structural half of the two-times problem: it shows that physical time, correctly described, is directed, inhomogeneous, structurally irreversible, and open toward the future, so that the temporal features tracked by biological systems are features of the physical world. To say what it is like to inhabit that structure from within lies outside the scope of the EBU, and the task of the EBU does not require it. Bergson's \emph{dur\'ee} names this first-person residue, and its location at the final step $T_B \to \textit{dur\'ee}$ of the hierarchy~\eqref{eq:hierarchy} allows the two-times problem to be stated with its physical and experiential halves kept distinct.

\section*{Statements and Declarations}
\noindent\textbf{Conflict of interest.} The author declares that there is no conflict of interest.

\bigskip

\noindent\textbf{Acknowledgements.} The author wishes to thank Francesco Nappo for many insightful discussions and valuable feedback, which significantly improved the manuscript.
\bibliography{reference_arxiv_1026}

\end{document}